\documentclass[prd, twocolumn, nofootinbib, floatfix]{revtex4-1}

\usepackage[mathscr]{euscript}
\usepackage{amsmath}
\DeclareMathOperator{\csch}{csch}

\usepackage{graphicx}
\usepackage{dcolumn}
\usepackage{bm}
\usepackage{epsfig}
\usepackage{amssymb,latexsym,mathrsfs}
\usepackage{graphicx}
\usepackage{color}
\usepackage{hyperref}
\usepackage{float}
\usepackage{diagbox}
\usepackage{physics}
\usepackage{braket}
\usepackage{tensor}

\definecolor{vividviolet}{rgb}{0.62, 0.0, 1.0}
\definecolor{amaranth}{rgb}{0.9, 0.17, 0.31}
\definecolor{palatinateblue}{rgb}{0.15, 0.23, 0.89}
\definecolor{brightpink}{rgb}{1.0, 0.0, 0.5}
\definecolor{cornflowerblue}{rgb}{0.39, 0.58, 0.93}
\definecolor{deepcarminepink}{rgb}{0.94, 0.19, 0.22}
\definecolor{radicalred}{rgb}{1.0, 0.21, 0.37}

\hypersetup{ linktoc=all,
    colorlinks, linkcolor={palatinateblue},
    citecolor={brightpink}, urlcolor={amaranth}
}

\newcommand{\D}{\mathrm{d}}
\newcommand{\be}{\begin{equation}}
\newcommand{\ee}{\end{equation}}
\newcommand{\bs}{\begin{split}} 
\newcommand{\bea}{\begin{eqnarray}}
\newcommand{\eea}{\end{eqnarray}}

\newcommand{\kp}{\kappa}

\renewcommand{\d}[1]{\ensuremath{\operatorname{d}\!{#1}}}

\newcommand{\bo}{\raise-1mm\hbox{\Large$\Box$}}

\newcommand{\f}[2]{\frac{#1}{#2}}

\newcommand{\w}{\omega}

\newcommand{\bes}{\begin{subequations}}
\newcommand{\ees}{\end{subequations}}

\begin{document}

\title{Relativistic quantum information as radiation reaction: entanglement entropy and self-force of a moving mirror analog to the CGHS black hole}
\author{Aizhan Myrzakul${}^{1}$}
\email{aizhan.myrzakul@nu.edu.kz}
\author{Chi Xiong${}^{2,3}$}
\email{xiong.chi@xmu.edu.my}
\author{Michael R.R. Good${}^{1}$}
\email{michael.good@nu.edu.kz}
\affiliation{${}^1$Physics Department \& Energetic Cosmos Laboratory, Nazarbayev University,\\ 53 Kabanbay Batyr, Nur-Sultan, 010000, Kazakhstan\\
${}^2$School of Mathematical and Physical Sciences, Nanyang Technological University,\\ 50 Nanyang Avenue, 639798, Singapore\\
${}^3$Department of Mathematics and Applied Mathematics, \\
Xiamen University Malaysia, \\
Selangor, 43900, Malaysia
}

\begin{abstract} 
The CGHS black hole has a spectrum and temperature that corresponds to an accelerated reflecting boundary condition in flat spacetime.  The  beta coefficients are identical to a moving mirror model where the acceleration is exponential in laboratory time.  The center and the event horizon of the black hole are at the same location modeled by the perfectly reflecting regularity condition that red-shifts the field modes.  In addition to computing the energy flux, we find the corresponding parameter associated with the black hole mass and the cosmological constant in the gravitational analog system. Generalized to any mirror trajectory we derive the self-force (Lorentz-Abraham-Dirac) and  express it and the power (Larmor) in connection with entanglement entropy, inviting an interpretation of acceleration radiation in terms of information flow. The mirror self-force and radiative power are applied to the particular CGHS black hole analog moving mirror which reveals the physics of information at the horizon during asymptotic approach to thermal equilibrium. 
\end{abstract} 

\date{\today} 

\maketitle

\section{Introduction} 
Three decades ago, several (1+1)-dimensional black hole models were introduced to gain insight into the quantum nature of black hole radiation, with one of the most prominent and physically interesting models being the Callan-Giddings-Harvey-Strominger (CGHS) system \cite{Callan:1992rs}. Simplified CGHS models, albeit with certain limitations, are exactly soluble and lead to many associated discoveries. New surprises related to complexity, temperatures, and entropy are still being found 
\cite{Strominger:1994tn,Giddings:1994pj,Thorlacius:1994ip,Fabbri,Ashtekar:2010hx}.

Moving mirrors are accelerated boundaries that create energy, particles, and entropy. They are simplified (1+1)-dimensional versions of the dynamical Casimir effect \cite{moore1970quantum,DeWitt:1975ys}. Interesting in their own right, they also act as toy models for black hole evaporation \cite{Hawking:1974sw,Unruh:1976db,Davies:1976hi,Davies:1977yv,Juarez-Aubry:2018ofz,Cong:2020nec,Wilson:2019ago}. 
The general and physically relevant connections of moving mirrors to black hole physics can be found in canonical textbooks \cite{Fabbri,Birrell:1982ix} and also in recent works, e.g. \cite{Good:2016atu, Myrzakul:2018bhy,Good:2018ell,Lin:2020itp}.

There have been a number of studies that relate different specific black hole models (e.g. the Schwarzschild \cite{Good:2016oey} case)  and their analog moving mirrors, including the extremal Reissner-Nordstr\"om \cite{Liberati:2000sq,good2020extreme}, extremal Kerr \cite{Rothman:2000mm}, Reissner-Nordstr\"om \cite{good2020particle}, Taub-NUT \cite{Foo:2020xmy} and Kerr \cite{Good:2020fjz} black holes. In addition, de Sitter and anti-de Sitter cosmologies \cite{Good:2020byh} are also modeled by moving mirror trajectories. For appropriately chosen trajectories  \cite{Good:2018aer}, close comparisons can be made with the radiation emitted from dynamic spacetimes \cite{wilczek1993quantum, Good:2017kjr}. Such an equivalence between a mirror and a curved spacetime is called an accelerated boundary correspondence (ABC).  

Our motivation in this paper is to synthesize and strongly link the well-known and important CGHS black hole model with its analog moving mirror counterpart.  In the process we want to derive the spectrum of particle production exactly and analytically, drawing close parallels between the two systems via the temperature, horizons and parameter analogs associated with CGHS black hole mass and cosmological constant. Furthermore, we aim to initiate an investigation into the entanglement entropy of a generalized mirror system and its relationship to the self-force on the mirror and power of the emitted vacuum radiation.  As we shall see, this link of inquiry reveals a close connection between the seemingly distinct concepts of self-force and information.  Application of the results for the CGHS mirror reveals the divergent self-force is directly a consequence of information loss. 

Additionally, we push the correspondence further, by considering the close connection to classical electrodynamic analogies.  The moving mirror is found to behave almost like a neutral particle coupled to the massless scalar field (similar to a charged particle coupled to the electromagnetic field).  Hence some of the familiar radiation results in classical electrodynamics has direct correspondence in the mirror case.  

The paper is organized in the following: in Sec.\ \ref{sec:actionEOM} we briefly review the CGHS action, the corresponding field equations of motion, and the formation of the CGHS black hole. In Sec.\ \ref{sec:CGHS metric}, the details of the CGHS metric and the transformation of this coordinate system to an accelerated mirror trajectory are investigated. In Sec.\ \ref{sec:expomirror}, we derive the particle flux radiated from the exponentially accelerated mirror in laboratory time and demonstrate its thermal characters for late-times. In Sec.\ \ref{sec:CGHS beta}, the particle flux radiated from the CGHS black hole is reviewed. Sec.\ \ref{sec:Quantum Stress tensor} and \ref{sec:Entropy} are dedicated to an overview of the quantum stress tensor and mirror entanglement entropy, respectively. In Sec.\ \ref{sec:power & force}, the moving mirror Larmor formula and Lorentz-Abraham-Dirac (LAD) force analogs are derived with an emphasis on entanglement entropy. In Sec.\ \ref{sec:CGHS power & force}, the formulas derived in the previous section are used to find the radiative power and radiation reaction force for our particular moving mirror (CGHS). In the last section, Sec. \ref{sec:summary}, the properties of the correspondence between the CGHS black hole and the exponentially accelerated mirror as well as the relation between the moving mirror and electrodynamics are summarized. In addition, some insight into future directions are provided. We use $\hbar = c =1$ except in the results of Eq. (\ref{power1}) and Eq. (\ref{force}) of Sec.\ \ref{sec:CGHS power & force}.

\section{Action \& Field Equation}\label{sec:actionEOM}
In this section we will briefly summarize the action and field equations of the Callan–Giddings–Harvey–Strominger model in which a linear dilaton vacuum evolves into a black hole from matter injection.  The CGHS action reads as,
\begin{equation} \label{CGHS}
S= \frac{1}{2} \int d^2x\sqrt{-g}\left[e^{-2\phi}\left(R+4(\nabla\phi)^2+4\Lambda^2\right)-|\nabla \chi|^2\right],    
\end{equation}
where $g$ is the metric tensor, $\phi$ is the dilaton field, $\Lambda$ is the cosmological constant, and $\chi$ are the matter fields.  
To obtain the equations of motion one may vary the action, Eq.~(\ref{CGHS}), with respect to the metric $g^{ab}$ and the dilaton field $\phi$, respectively,
\begin{eqnarray} \label{EOM}
2 e^{-2 \phi} \left[ \nabla_a \nabla_b \phi  + g_{ab} ( (\nabla \phi)^2 - \nabla^2 \phi - \Lambda^2) \right] &=& T_{ab}, \cr
e^{-2 \phi} \left[ - R +  4 (\nabla \phi)^2 -  4\nabla^2 \phi - 4 \Lambda^2) \right] &=& 0.
\end{eqnarray}
where $T_{ab} \equiv \nabla_a \chi \nabla_b \chi - \frac{1}{2} g_{ab} (\nabla \chi)^2$. 
Following \cite{Fabbri}, one can readily solve Eq.~(\ref{EOM}) in the conformal gauge, $ds^2= - e^{2\rho} dx^+ dx^-$. The solution is $\rho = \phi$ (gauge fixing) and
\begin{equation} 
  e^{-2\rho} = \frac{M(x^+)}{\Lambda} - \Lambda^2 x^+ \left(x^- + \frac{\mathcal{C}(x^+)}{\Lambda^2} \right),\label{solution}
\end{equation}
where the functions $M(x^+)$ and $\mathcal{C}(x^+)$ are integrals depending on the stress-energy tensor of the matter field, connected to the mass of CGHS black hole and the event horizon, respectively. Assume that we start with a linear dilaton vacuum, then turn on the matter flux injected to the system at some time $x^+_i$ and turn it off after the time $x^+_f$. Then when $x^+ > x^+_f$ the geometry of the system will approach and finally settle down to the static CGHS black hole background. The value $M(x^+_f)$ becomes the mass of the black hole and $\mathcal{C}(x^+_f)$ gives the curve of the event horizon. Therefore one can observe how the linear dilaton vacuum (before the time $x^+_i$) evolves into a CGHS black hole (after the time $x^+_f$) due to the matter injection.

\section{CGHS black hole and matching condition}\label{sec:CGHS metric}

In this section we concentrate on the CGHS black hole solution and some of the significant observable quantities within this model. The relevant metric for the CGHS black hole system can be cast in the form \cite{Giddings:2015uzr} (c.f. Schwarzschild gauge \cite{Fabbri}), 
\begin{equation}
ds^2=-f(r)dt^2+f(r)^{-1}dr^2,   \label{CGHSds}
\end{equation}
where 
\begin{equation}
f(r)=1-\frac{M}{\Lambda}e^{-2\Lambda r},    
\end{equation} 
with $\Lambda > 0$ as the cosmological constant parameterization scale of the spacetime and $M>0$ as the mass of the CGHS black hole. The curve of the event horizon function $\mathcal{C}(x^+)$ is set to zero in Eq.~(\ref{solution}) in order to obtain the metric for stationary CGHS black hole, Eq.~(\ref{CGHSds}). Notice that when $M=\Lambda$, Eq.\ (\ref{CGHSds}) possesses a singularity at the radial coordinate $r = 0$ which also reduces to the event horizon, in contrast to the coordinate singularity of the Schwarzschild black hole event horizon (see e.g. the Schwarzschild mirror \cite{Good:2016oey,Good_2017Reflections,Anderson_2017,Good_2017BHII}).  For general $M$ and $\Lambda$, the horizon is at $r_H = \frac{1}{2\Lambda} \ln \frac{M}{\Lambda}$. The surface gravity of the CGHS black hole can be obtained as \cite{derek2009black},
\begin{align}
    \kappa &= \frac{1}{2} \frac{\D}{\D r}f(r) \Big|_{r = r_H} = \Lambda,
    \end{align} 
where $r_H = 0$, so that consistency with the laws of black hole thermodynamics dictates the temperature is $T = \Lambda/(2\pi)$. For a double null coordinate system ($u,v$) with $u = t-r^\star$ and $v = t + r^\star$, the associated tortoise coordinate $r^\star$ can be obtained in the usual way \cite{Fabbri}, via
\begin{align}
    r^\star &= \int \frac{\D r }{f(r)},\label{findtort}
\end{align}
which yields 
\be
r^*=\frac{1}{2\Lambda} \ln \left|\frac{M}{\Lambda}-e^{2 \Lambda  r}\right|. \label{tortoise1}   
\ee
The absolute brackets are critical for real coordinate values.  

Following Wilczek \cite{wilczek1993quantum}, let us coincide the inner and outer regions of a collapsing null shell to form a black hole, where the exterior background is given by the CGHS metric,
\begin{equation}
  ds^2=\begin{cases}
    -d t_{in}^2+dr^2, & \text{for $t_{in}+r\leq v_0$},\\
    -fdt_{out}^2+f^{-1}dr^2, & \text{for $t_{out}+r\geq v_0$}.
  \end{cases}\label{Matching}
\end{equation}
and $v_0$ is a light-like shell.
In null coordinates, the system Eq.~(\ref{Matching}) can be rewritten as, 
\begin{equation}
  ds^2=\begin{cases}
    -dUdV, & \text{where $U=t_{in}-r$, $V=t_{in}+r$},\\
    -fdudv, & \text{where $u=t_{out}-r_*$, $v=t_{out}+r_*$}.\label{matching2}
  \end{cases}
\end{equation}
So, the metric for the geometry describing the outside region of a collapsing shell takes the simplified form, $\D s^2 = -f d u d v $. 

The matching condition (see \cite{Fabbri,wilczek1993quantum,Good:2020fjz}) with the flat interior geometry, described by the interior coordinates $(U,V)$ is the trajectory corresponding to $r = 0$, expressed in terms of the exterior function $u(U)$. We can obtain this matching via the association $r = r^\star$, and taking $r^\star ( r = (v_0 - U) /2 ) = (v_0 - u )/2$ along the light ray, $v_0$. We set $v_0=0$ for simplicity without loss of generality. This matching condition,
\be
u(U) = -\frac{1}{\Lambda} \ln \left|\frac{M}{\Lambda}-e^{-\Lambda U}\right|,
\ee
is the outside $u$ trajectory of the origin as a function of the inside coordinate $U$.  We can write this as,
\be
u(U) = U -\frac{1}{\Lambda} \ln \left(1-\frac{M}{\Lambda} e^{\Lambda U}\right) \label{CGHSmatch},
\ee
where $U<0$ and $\Lambda >0$.  The regularity condition of the modes requires that they vanish at $r = 0$, which acts as a reflecting boundary in the black hole system. In the accelerated boundary correspondence (ABC) of the mirror system, the origin of the black hole functions as the mirror trajectory in flat spacetime.  The position of the origin is a dynamic function $u$ with independent variable $U$.  Since the field vanishes (does not exist for $r<0$), the form of the field modes can be determined, allowing for the identification $U\Leftrightarrow v$ (where $v$ is the flat spacetime advanced time in the moving mirror model) for the Doppler-shifted field modes. In the next section we will define the analog mirror trajectory for the CGHS spacetime by making the identification $u(U) \Leftrightarrow f(v)$, which is a known function of the advanced time $v$.

\section{Exponentially Accelerated Mirror}\label{sec:expomirror}

In this section we focus on the trajectory and particle flux radiation of the exponentially accelerated mirror in coordinate time to demonstrate their equivalence with the corresponding quantities in the CGHS black hole model. 

\begin{figure}[h]
    \centering
    \includegraphics[width=0.9\linewidth]{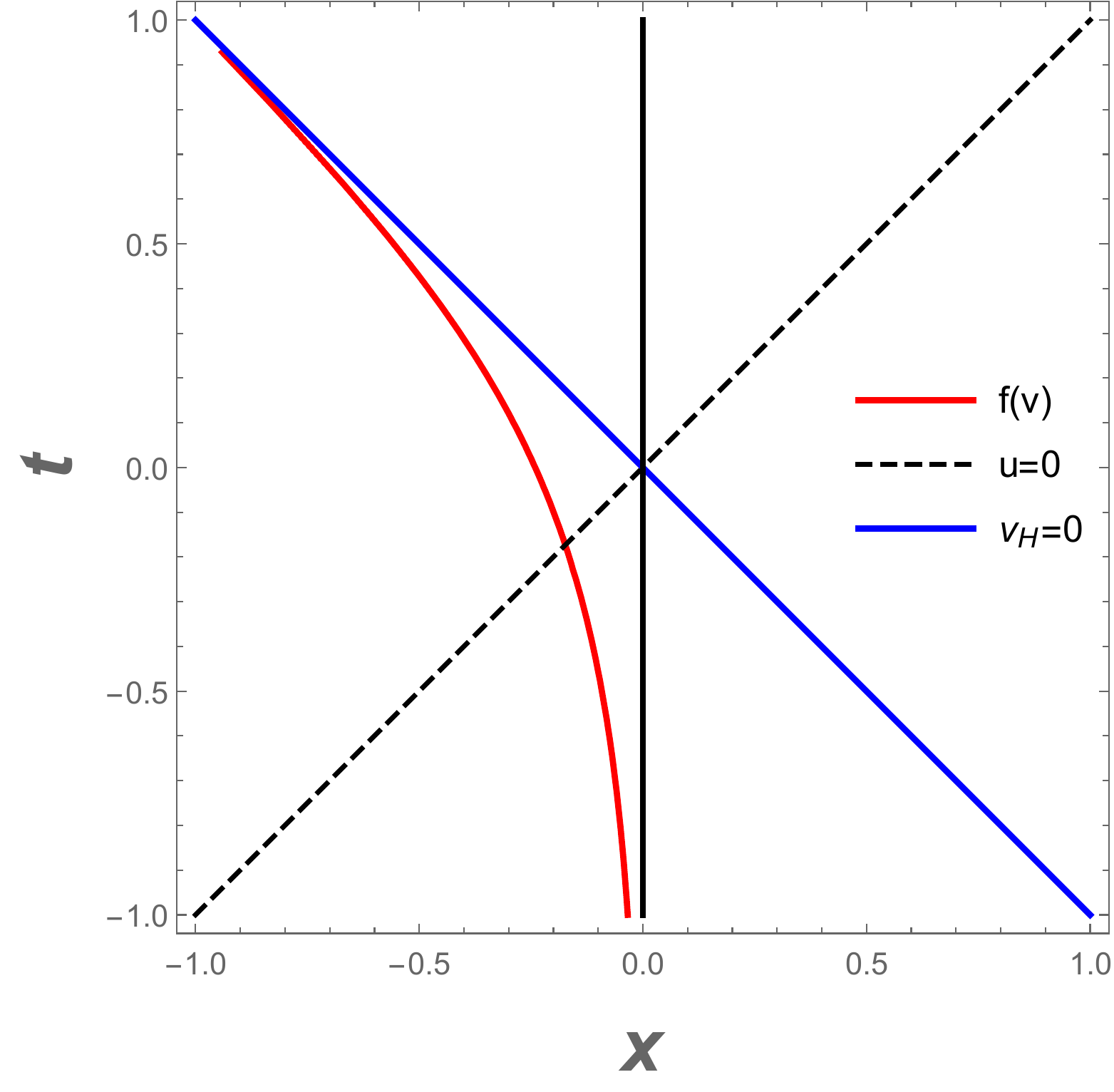}
    \caption{Spacetime diagram with $M=\Lambda =2$, of Eq.~(\ref{f(v)}), $f(v)$.  The mirror starts asymptotically timelike and finishes asymptotically lightlike with infinite acceleration along the horizon $v_H=0$.    } 
    \label{fig:spacetime}
\end{figure}

\begin{figure}[h]
    \centering
    \includegraphics[width=0.9\linewidth]{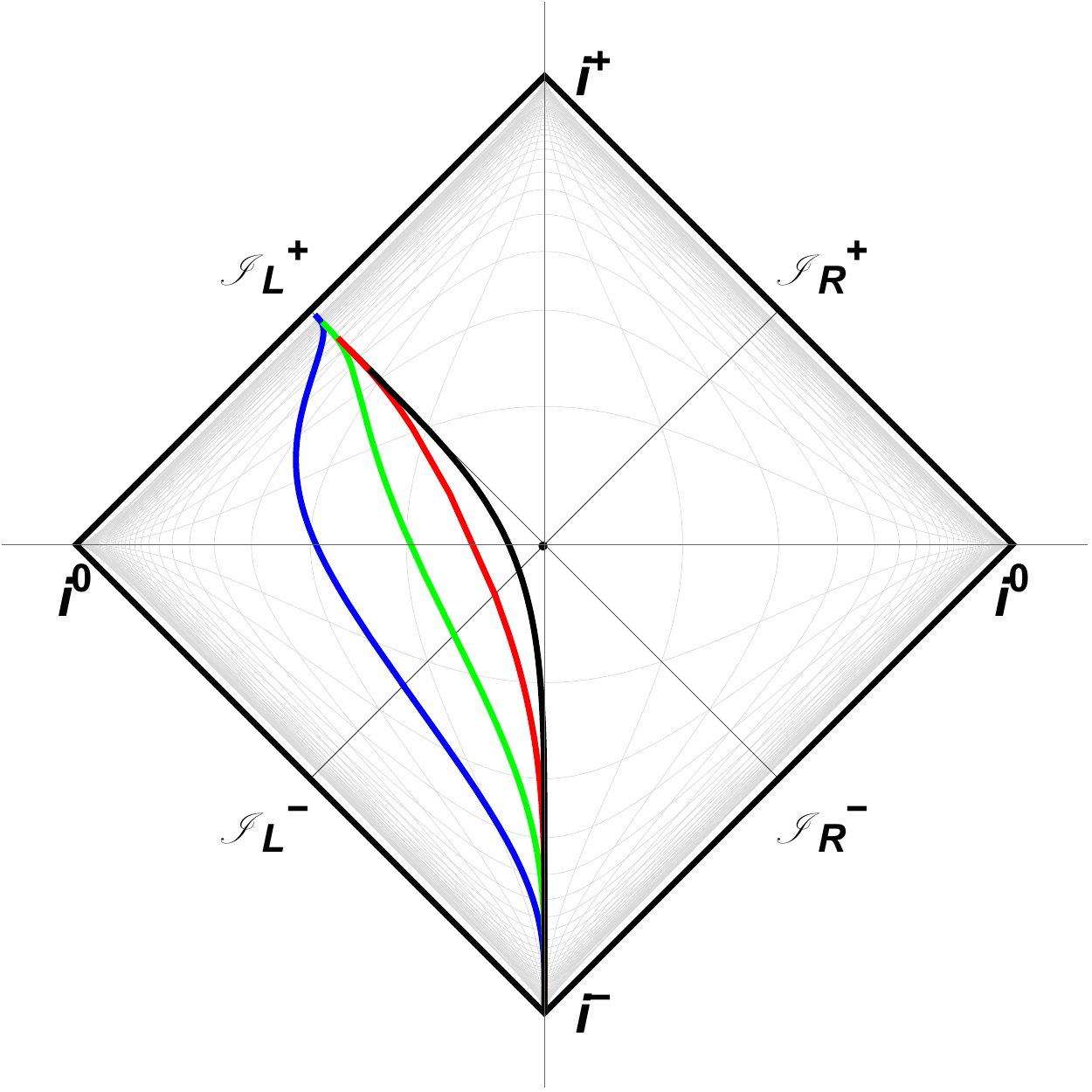}
    \caption{Penrose diagram with Blue, Green, Red, Black, parametrized respectively with $\kappa = 1/2, 1, 2, 4$ of Eq.~(\ref{f(v)}), $f(v)$.  As in Figure \ref{fig:spacetime} (the Red line is the same trajectory) the mirror starts asymptotically timelike and finishes asymptotically lightlike with infinite acceleration along the horizon $v_H=0$.   } 
    \label{fig:penrose}
\end{figure}

In line with previous accelerated boundary correspondences, consider the exponentially accelerated mirror trajectory with proper acceleration \cite{Good:2017ddq}: 
\be \alpha(t) = -\f{\kp}{2} e^{\kp (t-v_H)}, \ee
where $\kappa > 0$ is a parameter of the acceleration and $v_H$ is the horizon in advanced time, $v = t+x$. This $x$ and $t$ are the usual lab coordinates of flat (1+1)-dimensional Minkowski spacetime. 
The trajectory in light cone coordinates as a function of advanced time is
\begin{equation}
f(v)=v-\frac{1}{\kappa}\ln\left(1-e^{\kappa(v-v_H)}\right), \label{f(v)}   
\end{equation} where, identifying Eq.~(\ref{f(v)}) with Eq.~(\ref{CGHSmatch}) as usual (see prior ABCs), the associated parameters in the CGHS system define the moving mirror's null-ray horizon, 
\begin{equation}
v_H=\frac{1}{\Lambda}\ln\left(\frac{\Lambda}{M}\right)\label{Arcx_horizon},
\end{equation}
which is the location that the last incoming left-moving ray reflects off the mirror.  Past this position, there is no more reflection and left-mover modes never make it to an observer at right null-infinity, $\mathscr{I}^+_R$. The mirror horizon couples the parameters $\Lambda$ and $M$, which are the cosmological constant and mass of the black hole, respectively, in the CGHS system. The fact that $v_H$, which is the finite $v$ for the mirror horizon, is also closely related to the CGHS black hole horizon, through $2r_H = -v_H$, further corroborates a correspondence between the CGHS black hole and the exponentially accelerated mirror.
A spacetime plot of this asymptotic light-like moving mirror is given in Fig. \ref{fig:spacetime}.  A Penrose conformal diagram is given in Fig. \ref{fig:penrose}.  Notice when $v_H=0$, Eq.~(\ref{f(v)}) is  Eq.~(\ref{CGHSmatch}), i.e. $u(U)\Leftrightarrow f(v)$, when $\Lambda = M$.  

It is interesting to note that in Schwarzschild black hole case the singularity is located at the center, that corresponds to the mirror, and the event horizon is located at $r=2M$. Unlike the Schwarzschild black hole, for the CGHS black hole the singularity happens at $r=0$ which is the location of event horizon as well when $\Lambda = M$. So, the mirror mimics both the event horizon and the center of the black hole, simplifying physical interpretation and giving a straightforward answer to the origin of particle creation in the CGHS system. 

Now we will derive the thermal Planck distribution of the exponentially accelerated moving mirror particles by use of the beta Bogolubov coefficient. The beta coefficient can be found via an integration \cite{Good:2016atu} by parts where we ignore non-contributing surface terms, 
\be\label{betap} \beta_{\w\w'} = \frac{1}{2\pi}\sqrt{\frac{\w'}{\w}}\int_{-\infty}^{v_H =0} dv \; e^{-i\w' v}e^{-i\w f(v)},  \ee
to get
\begin{equation}
\beta_{\omega\omega'}=\frac{1}{2\pi\kappa}\sqrt{\frac{\omega'}{\omega}}B\left[-\frac{i\w_+}{\kappa},1+\frac{i\omega}{\kappa}\right], \label{beta_final}   
\end{equation}
where we utilize the Euler integral of the first kind as a Beta function, $B(a,b) = \f{\Gamma(a)\Gamma(b)}{\Gamma(a+b)}$, and $\omega_+ = \omega + \omega'$. 
Multiplying by its complex conjugate gives the particle count per $\w'$ mode, per $\w$ mode:
\begin{equation}
|\beta_{\omega\omega'}|^2=\frac{1}{4\pi^2\kappa^2}\frac{\omega'}{\omega}\left|B\left[\frac{i\omega_+}{\kappa},1-\frac{i\omega}{\kappa}\right]\right|^2, \label{Arcx_beta} 
\end{equation}
or, equivalently expressed,
\begin{equation}
|\beta_{\omega\omega'}|^2 = \frac{e^{\frac{2 \pi  \omega }{\kappa }} \left(e^{\frac{2 \pi  \omega '}{\kappa }}-1\right)}{2 \pi  \kappa \omega_+  \left(e^{\frac{2 \pi  \omega }{\kappa }}-1\right) \left(e^{\frac{2 \pi  \omega_+}{\kappa }}-1\right)}. 
\end{equation}
Thermal character results in the high frequency limit $\w'\gg \w$ approximation (a good explanation for how this corresponds to late times is given by Hawking \cite{Hawking:1974sw}).  We can see by inspection that,
\be |\beta_{\w\w'}|^2 \approx \frac{1}{2\pi\kappa \omega'} \f{1}{e^{2\pi \w/\kp} - 1} \quad \text{for}\; \w'\gg\w \ee 
so that $ T = \kappa/(2\pi)$ at late times.

\section{CGHS particle radiation}\label{sec:CGHS beta}

In this section we calculate the particle flux radiated by the CGHS black hole and the corresponding beta Bogolubov coefficients, following the standard procedure (see e.g. \cite{Fabbri}). Remarkably, the beta Bogolubov coefficients match those corresponding ones obtained from the particle radiation of the exponentially accelerating mirror (Section \ref{sec:expomirror}).

The standard Bogolubov procedure for calculating the Hawking radiation considers two relevant regions as mentioned in Sec.\ \ref{sec:actionEOM}: the linear dilaton vacuum and the CGHS black hole, described by the corresponding ``in" and ``out" coordinates which can be connected via the Kruskal coordinates (see \cite{Fabbri} for details and elaboration on the standard notation). One then identifies plane wave modes for ingoing and outgoing sectors using null Minkowski coordinates $(\sigma_{in}^\pm,\sigma_{out}^\pm)$. They are related as,
\begin{equation}
\sigma_{in}^+=\sigma_{out}^+=\sigma^+,  \label{sigma_in}
\end{equation}
\begin{equation}
\sigma_{in}^-=-\frac{1}{\Lambda}\ln\left(e^{-\Lambda\sigma_{out}^-}+\frac{\mathcal{C}(x_f^+)}{\Lambda}\right). \label{sigma_out}  
\end{equation}
So, using $\sigma^-$ sector the plane wave modes have the following forms,
\begin{equation}
g_{\omega'}^{in}=\frac{1}{\sqrt{4\pi\omega'}}e^{-i\omega'\sigma_{in}^-}, \label{mode_in}   
\end{equation}
\begin{equation}
g_\omega^{out}=\frac{1}{\sqrt{4\pi\omega}}e^{-i\omega\sigma_{out}^-},\label{mode_out}
\end{equation}
designated by $g$ (sometimes $u$ is used but here $u$ is already a retarded time null coordinate). The next step is to evaluate the beta Bogolubov coefficients by calculating scalar product between the plane wave modes. For our particular case the corresponding integral is,
\begin{equation}
\beta_{\omega\omega'}=(g_\omega^{out},g_{\omega'}^{in*})=2i\int_{-\infty}^{+\infty}d\sigma_{in}^-g_\omega^{out}\frac{\partial g_{\omega'}^{in}}{\partial\sigma_{in}^-}. \label{scalar_product}   
\end{equation} 
Substituting Eq.~(\ref{mode_in}) and Eq.~(\ref{mode_out}) into Eq.~(\ref{scalar_product}), and also using Eq.~(\ref{sigma_out}), the above integral becomes,
\begin{equation}
\beta_{\omega\omega'}=\frac{1}{2\pi}\sqrt{\frac{\omega'}{\omega}}\int_{-\infty}^{\sigma_{in,H}^-}d\sigma_{in}^{-}\frac{e^{\frac{i\omega}{\Lambda}\ln\left(e^{-\Lambda\sigma_{in}^{-}-\frac{\mathcal{C}(x_f^+)}{\Lambda}}\right)}}{e^{i\omega'\sigma_{in}^{-}}},   \label{beta1}
\end{equation}
where
\begin{equation}
\sigma_{in,H}^-=\frac{1}{\Lambda}\ln\frac{\Lambda}{\mathcal{C}(x_f^+)}\label{CGHS_horizon}
\end{equation}
is the black hole event horizon location that is formed when injecting matter into linear dilaton vacuum. 
Note that Eq. (\ref{CGHS_horizon}) is similar to Eq. (\ref{Arcx_horizon}), i.e. $v_H\Leftrightarrow\sigma_{in,H}^-$. 
Calculation of Eq.~(\ref{beta1}) gives, 
\begin{equation}
\beta_{\omega\omega'}=\frac{1}{2\pi\Lambda}\sqrt{\frac{\omega'}{\omega}}\left(\frac{\mathcal{C}(x_f^+)}{\Lambda}\right)^{\frac{i\omega_+}{\Lambda}}B\left[-\frac{i\omega_+}{\Lambda},1+\frac{i\omega}{\Lambda}\right]. \label{CGHS_beta}   
\end{equation}
The complex conjugate squared of Eq. (\ref{CGHS_beta}) yields,
\begin{equation}
|\beta_{\omega\omega'}|^2=\frac{1}{4\pi^2\Lambda^2}\frac{\omega'}{\omega}\left|B\left[\frac{i\omega_+}{\Lambda},1-\frac{i\omega}{\Lambda}\right]\right|^2,   \label{CGHS_betasquare} \end{equation}
which is exactly Eq. (\ref{Arcx_beta}) given that $|\Lambda|=|\kappa|$. 
Interestingly, the mass of the CGHS black hole, $M(x^+_f)$, does not appear in  $\beta_{\omega\omega'}$, and $\mathcal{C}(x^+_f)$ disappears when calculating  $|\beta_{\omega\omega'}|^2$. This is because the spectrum of the CGHS black hole does not explicitly depend on its mass. The correspondence between the black hole mass and the curve of the event horizon, $\mathcal{C}(x_f^+)$, can be seen from the comparison of Eq. (\ref{Arcx_horizon}) and Eq. (\ref{CGHS_horizon}). The curve of the event horizon is defined by the so-called ``apparent horizon", $x^- = - \mathcal{C}(x^+)/\Lambda^2$, which is spacelike or null and coincides with the event horizon after the matter injection has finished at time $x^+_f$, i.e. after time $x^+_f$ when the geometry of the CGHS black hole is settled and the apparent horizon becomes the event horizon. 

Hereinafter, due to the identical particle production Eq.~(\ref{Arcx_beta}) between the exponentially accelerated mirror in coordinate laboratory time and the CGHS system Eq.~(\ref{CGHS_betasquare}), for short, we refer to this specific perfectly reflecting boundary trajectory, Eq.~(\ref{f(v)}), as the CGHS mirror.

\section{Quantum stress-tensor}\label{sec:Quantum Stress tensor}
In this section we will briefly review the quantum stress tensor of the moving mirror following Davies and Fulling \cite{Davies:1976hi,Davies:1977yv}. We will need this to specialize to the CGHS mirror and compute its energy flux which we do in Section \ref{sec:CGHS power & force}.  In (1+1)-dimensional flat spacetime the energy-momentum tensor is determined by the following $2\times 2$ matrix,
\begin{equation}
T_{\mu\nu}=\frac{1}{2}\begin{bmatrix}
\left(\frac{\partial\phi}{\partial t}\right)^2+\left(\frac{\partial\phi}{\partial x}\right)^2 & \frac{\partial\phi}{\partial x}\frac{\partial\phi}{\partial t}+\frac{\partial\phi}{\partial t}\frac{\partial\phi}{\partial x} \\
\frac{\partial\phi}{\partial t}\frac{\partial\phi}{\partial x}+\frac{\partial\phi}{\partial x}\frac{\partial\phi}{\partial t} & \left(\frac{\partial\phi}{\partial t}\right)^2+\left(\frac{\partial\phi}{\partial x}\right)^2
\end{bmatrix}.  \label{tensor_matrix}  
\end{equation}
Here $\phi(t,x)$ is technically a free field that obeys the massless scalar wave equation,
\begin{equation}
\frac{\partial^2\phi}{\partial t^2}-\frac{\partial^2\phi}{\partial x^2}=0,
\end{equation}
but there are effective interactions with the boundary condition,
\begin{equation}
\phi(t,x)|_{z}=0,    
\end{equation} which is imposed on the field equation of motion, where $x=z(t)$ is the trajectory of the moving mirror. In this two dimensional case, the mirror is a perfectly reflecting point moving along a timelike worldline $z(t)$.

In the usual quantum field theory, $\phi(t,x)$ is an operator defined by field modes as,
\begin{equation}
\phi(t,x)=\int_0^\infty\left[\hat{a}_\omega^{in}\phi_\omega+\hat{a}_\omega^{in\dagger}\phi_\omega^*\right]d\omega,  \label{field_operator}  
\end{equation}
where $\hat{a}_\omega$ and $\hat{a}_\omega^\dagger$ are ladder operators, and $\phi_\omega^*$ is a complex conjugate of $\phi$. After inserting Eq. (\ref{field_operator}) into the matrix Eq.~(\ref{tensor_matrix}), the stress-tensor can be written as,
\begin{equation}
T_{\mu\nu}=:T_{\mu\nu}:+\braket{T_{\mu\nu}}, \label{stress-tensor}   
\end{equation}
where the first term has normal ordering of the ladder operators, i.e.
\begin{equation}
:\hat{a}_\omega^{in}\hat{a}_\omega^{in\dagger}:=\hat{a}_\omega^{in\dagger}\hat{a}_\omega^{in}.   
\end{equation}
The second term in Eq. (\ref{stress-tensor}) is the expectation value of the operator in vacuum, which is defined as,
\begin{equation}
\braket{T_{\mu\nu}}=\int_0^\infty  T_{\mu\nu}(\phi_\omega,\phi_\omega^*)d\omega. \label{expectation_value}  
\end{equation} The stress tensor here, as it stands, is of significant interest, but unfortunately the above integral is divergent. In order to make the integral in Eq.~(\ref{expectation_value}) finite and extract useful information about the emitted radiation, point-splitting regularization is utilized.  The key idea is to evaluate the field modes at different times:  
$\phi$ at $(t,x)$ and $\phi^*$ at $(t+\epsilon,x)$, respectively, where $\epsilon$ is an infinitesimally small quantity. So, the field modes at corresponding points are:
\begin{equation}
\begin{cases}
\begin{matrix}
\frac{\partial\phi_\omega}{\partial t}\\
\frac{\partial\phi_\omega}{\partial x}
\end{matrix}
\end{cases}=
\sqrt{\frac{\omega}{4\pi}}\left[e^{-i\omega v}\mp p'(u)e^{-i\omega p(u)}\right],\label{mode}
\end{equation}
\begin{equation}
\begin{cases}
\begin{matrix}
\frac{\partial\phi_\omega^*}{\partial t}\\
\frac{\partial\phi_\omega^*}{\partial x}
\end{matrix}
\end{cases}=
\sqrt{\frac{\omega}{4\pi}}\left[e^{i\omega (v+\epsilon)}\mp p'(u+\epsilon)e^{i\omega p(u+\epsilon)}\right],    \label{mode_conjugate}
\end{equation}
where $p(u)\equiv2t(u)-u$ and $u\equiv t-z(t)$. Inserting Eqs. (\ref{mode}) and (\ref{mode_conjugate}) into Eq. (\ref{expectation_value}) leads to,
\begin{equation}
\begin{cases}
\begin{matrix}
 \braket{T_{00}}=\braket{T_{11}}\\
 \braket{T_{10}}=\braket{T_{01}}
 \end{matrix}=
 \frac{1}{4\pi}\int_0^\infty\omega\left[e^{i\omega\epsilon}\pm\frac{p'(u)p'(u+\epsilon)}{e^{-i\omega(p(u+\epsilon)-p(u))}}\right] d\omega .
 \end{cases}
\end{equation}
Calculation of the above integrals results in,
\begin{equation}
\begin{cases}
\begin{matrix}
 \braket{T_{00}}\\
 \braket{T_{01}}
 \end{matrix}=-\frac{1}{4\pi\epsilon^2}\mp\frac{1}{4\pi}\frac{p'(u)p'(u+\epsilon)}{[p(u)-p(u+\epsilon)]^2}.
 \end{cases}\label{tensors}
\end{equation}
Consequently,
\begin{equation}
\braket{T_{00}}=-\frac{1}{2\pi\epsilon^2}-\braket{T_{01}}.   \label{tensor} 
\end{equation}
The result for $\braket{T_{01}}$ in Eq.~(\ref{tensors}) is Taylor expanded in $\epsilon$ to give,
\begin{equation}
\braket{T_{01}}=\frac{1}{24\pi}\left[\frac{p'''}{p'}-\frac{3}{2}\left(\frac{p''}{p'}\right)^2\right]+O(\epsilon).   
\end{equation}
In the limit $\epsilon\rightarrow 0$, the first term in Eq. (\ref{tensor}) becomes divergent. Since the second term is independent of $\epsilon$, it does not vanish like the higher order terms. This second term is of particular physical interest, corresponding to the energy flux radiated by the mirror, 
\begin{equation}
\braket{T_{00}}=\mathcal{F}(u)= -\frac{1}{24\pi}\left[\frac{p'''}{p'}-\frac{3}{2}\left(\frac{p''}{p'}\right)^2\right].   
\end{equation}
The first component of the renormalized stress tensor expectation value gives energy flux radiated by the mirror into the vacuum, characterizing the amplified quantum fluctuations due to the presence of the accelerating boundary. 

Using the relations between spacetime $(t,x)$ and null $(u,v)$ coordinates as,
\begin{equation}
u\equiv t-z(t), ~~~\quad v\equiv t+z(t),
\end{equation}
and 
\begin{equation}
p(u)=2t(u)-u,~~~\quad f(v)=2t(v)-v,
\end{equation} the energy flux can also be expressed with care, using straightforward differential algebra,
\begin{equation}
\mathcal{F}(t)=\dfrac{\dddot{z}(\dot{z}^{2}-1)-3\dot{z}\ddot{z}^{2}}{12\pi(\dot{z}+1)^{2}(\dot{z}-1)^{4}},
\end{equation}
\begin{equation}
\mathcal{F}(x)=\dfrac{t'''(t'^{2}-1)-3t't''^{2}}{12\pi(t'+1)^{2}(t'-1)^{4}},\label{flux(x)}
\end{equation}
\begin{equation}
\mathcal{F}(v)=\dfrac{1}{24\pi}\bigg[\dfrac{f'''}{f'}-\dfrac{3}{2}\bigg(\dfrac{f''}{f'}\bigg)^{2}\bigg]\dfrac{1}{f'^{2}},
\end{equation}
where dots and primes denote derivatives with respect to arguments $t$, $x$ and $v$, respectively. Eq.~(\ref{flux(x)}) will be used to find the flux of the CGHS moving mirror in Sec.~\ref{sec:CGHS power & force}.

\section{Mirror Entanglement Entropy}\label{sec:Entropy}

In this section we will review the derivation of (1+1)-dimensional entanglement (geometric) entropy in conformal field theory (CFT) and its connection to the rapidity of the moving mirror (see other derivations e.g. \cite{Good:2020nmz, Fitkevich:2020okl}). 

Consider the entropy of a system in (1+1)-D CFT \cite{Holzhey:1994we}, 
\begin{equation}
S=\frac{1}{6}\ln\frac{L}{\epsilon}, \label{conformal_entropy}   
\end{equation}
where $L$ is the size of the system in general (and in our case it is the mirror trajectory which measures the size of the system by the spacetime traversed accessible to the quantum field), and $\epsilon$ is a UV cut-off.

For a general arbitrary moving mirror,
\begin{equation}
L\equiv p(u)-p(u_0),  \label{general_size}  
\end{equation}
where $u$ and $u_0$ are null coordinates that form the region in the system which we are considering, and $\epsilon$ is asymmetrically smeared, i.e. $\epsilon^2\equiv\epsilon_p\epsilon_{p_0}$. 
Here $p(u)$ is the trajectory of the mirror in null coordinates (it is a function of retarded time $u$).  The smearing and dynamics of the mirror are related as,
\be
\epsilon_p = p'(u)\epsilon_u, \qquad
\epsilon_{p_0} = p'(u_0)\epsilon_{u_0}.\label{p'(u)}\ee
Substituting Eqs. (\ref{p'(u)}) into Eq.~(\ref{conformal_entropy}) yields the bare entropy of the system,
\begin{equation}
S_{bare}=\frac{1}{12}\ln\frac{\left[p(u)-p(u_0)\right]^2}{p'(u)p'(u_0)\epsilon_u\epsilon_{u_0}}.    
\end{equation}
The vacuum entropy of the system can be found by considering a static mirror  where 
$L=u-u_0$ and $\epsilon^2=\epsilon_u\epsilon_{u_0}$. Thus,
\begin{equation}
S_{vac}=\frac{1}{12}\ln\frac{(u-u_0)^2}{\epsilon_u\epsilon_{u_0}}.    
\end{equation}
Even though the entropies above are defined in terms of smearing, this dependence can be removed by an intuitive renormalization via,
\begin{equation}
S_{ren}=S_{bare}-S_{vac}=\frac{1}{12}\ln\frac{[p(u)-p(u_0)]^2}{p'(u)p'(u_0)(u-u_0)^2}. \label{renorm_entropy}   
\end{equation}
Further simplification proceeds by a Taylor expansion of our arbitrary function $p(u)$ around $u=u_0$ up to first order, that is,
\begin{equation}
p(u)=p(u_0)+p'(u_0)(u-u_0)+\mathcal{O}(u-u_0)^2.  \label{expansion}  
\end{equation} Substituting Eq.~(\ref{expansion}) into Eq.~(\ref{renorm_entropy}) brings us to, 
\begin{equation}
S_{ren}=\frac{1}{12}\ln\frac{p'(u_0)}{p'(u)}.\label{renorm_entropy2}
\end{equation}
Moreover, for a static mirror $p(u)=u$ and $p(u_0)=u_0$, therefore $p'(u_0)=1$.
As a result, Eq. (\ref{renorm_entropy2}) reduces to $S_{ren}\rightarrow S(u)$ where,
\begin{equation}
S(u)=-\frac{1}{12}\ln p'(u).    \label{S(u)}
\end{equation}
Eq.~(\ref{S(u)}) is valid for any moving mirror that starts asymptotically static (zero velocity). Notable exceptions are the eternally thermal Carlitz-Willey mirror \cite{carlitz1987reflections} and the eternally uniformly accelerated mirror \cite{Birrell:1982ix}; however, most of the solved mirrors in the literature, by construction, do start static as they are often used to model gravitational collapse.  The CGHS mirror is no exception.

Eq.~(\ref{S(u)}) is more intuitively written in spacetime coordinates using the relation between null and spacetime trajectories of the mirror as,
\begin{equation}
p'(u)=\frac{1+\dot{z}(t)}{1-\dot{z}(t)}.   
\end{equation} Applying this relation into the Eq.~(\ref{S(u)}) yields,
\begin{equation}
S(t)=-\frac{1}{6}\tanh^{-1}[\dot{z}(t)]=-\frac{1}{6}\eta(t),\label{S(t)}
\end{equation}
where $\eta(t)\equiv\tanh^{-1}[\dot{z}(t)]$ is the time-dependent rapidity. It is simple to see that the magnitude of the entropy increases as the mirror moves faster. Unitarity in this context \cite{Bianchi2014} strictly requires that the entropy must achieve a constant value in the far past and far future. 

This von Neumann entropy measure of the degree of quantum entanglement between the two subsystems (past \& future) constitutes a two-part composite quantum system.  It explicitly reveals the connection of information of entanglement to the dynamics (rapidity) of the moving mirror system.  Allow us to speculate a thermodynamic treatment of the system, and the corresponding macroscopic state of the entanglement is characterized by a distribution of its microstates, then it may be appropriate to coincide the Boltzmann entropy with the von Neumann entropy. In this conjectural case, a discrete speed based on the basic smallest unit of an operable binary digit of information results.\footnote{This discreteness necessarily leads to a smallest non-zero speed for the moving mirror.  In SI units, this is $17.2 \;\textrm{fm/s}$, from $v = c \tanh(6 k_B \ln 2)$ which is about the diameter of a gold nucleus in one second, or about 5 centimeters in 100,000 years.} 

In the next section we will ultimately apply the above entanglement-rapidity relationship, Eq.~(\ref{S(t)}), which can be expressed independently of coordinates or its argument, $-6S=\eta$, to gain insight into the self-force and Larmor power by reformulating them in terms of entropy.

\section{Mirror Larmor formula and LAD force}\label{sec:power & force}

\subsection{Quantum relativistic Larmor formula}\label{power}
In this subsection we derive the quantum relativistic power radiated by the moving mirror and find it has the same form as the classical relativistic Larmor formula of electrodynamics.  We account for the power radiated to both sides of the moving mirror, utilizing the quantum stress tensor derived in Section \ref{sec:Quantum Stress tensor}.

Let us consider the total energy radiated as derived from the Davies-Fulling quantum stress tensor, to the right side of the mirror, expressed as Equation (2.34) of \cite{good2013time},
\be E^R = \frac{1}{12\pi} \int_{-\infty}^{\infty} \alpha^2(1+\dot{z})\,\d t.\ee
An observer at $\mathscr{I}^+_R$ measures $E^R$ energy emitted, but the mirror also radiates energy to $\mathscr{I}^+_L$ for an observer on the left. The energy radiated to the left, $E^L$, is found by the same expression but with a parity flip, $\dot{z} \rightarrow -\dot{z}$, so that
the total radiated energy is,
\be E = E^R+E^L =  \frac{1}{6\pi} \int_{-\infty}^{\infty} \alpha^2\,\d t.\ee
We define the quantity, $E$, without an average giving us a measure of the total radiation, independent of observer.  This allows us to identify and define a quantum relativistic Larmor power analog for the moving mirror, $P = \d E/\d t$,
\be E = \int_{-\infty}^{\infty} \frac{d E}{d t}\d t \equiv \int_{-\infty}^{\infty} P \d t,\ee
which gives the familiar relativistic Larmor scaling for proper acceleration:
\be P = \frac{\alpha^2}{6\pi}.\label{Larmor_power}
\ee
The quantum power radiated by the mirror takes the same form as that of a classical point charge in electrodynamics (see \cite{3D} for the derivation and distribution in (3+1) dimensions).  Recall here that $\alpha$ is the scalar invariant which is defined as the proper time derivative of rapidity, $\alpha = \eta'(\tau)$, even though the integral in which we define the power with respect to uses ordinary coordinate time. Eq.~(\ref{Larmor_power}) is in harmony with Ford-Vilenkin \cite{Ford:1982ct} who found that the self-force of the moving mirror also has the same form as radiation-reaction of a point charge in classical electrodynamics.

\subsection{Relativistic entanglement-power}\label{entanglementpower}
In this subsection we apply the entanglement-rapidity relationship, Eq.~(\ref{S(t)}), to the power derived in the previous subsection, Eq.~(\ref{Larmor_power}).  Motivated to understand the radiated emission in terms of information flow, we find the quantum power is expressed as the square of the first derivative of the entanglement.  

The covariant Larmor power $P$ is a Lorentz scalar invariant since it is proportional to the square of the proper acceleration, 
\be P = \frac{\alpha^2}{6\pi} = \frac{\eta'(\tau)^2}{6\pi}.\label{rp}\ee
The rapidity-entanglement relationship, given by $\eta = -6S$ (see also Section \ref{sec:Entropy} and further references \cite{Good:2016atu,Bianchi:2014qua,good2020extreme}), can be expressed with independent variable as proper time, $\eta'(\tau) = -6S'(\tau)$, and thus the power is formulated as,
\be P = \frac{6}{\pi} S'(\tau)^2.\label{EP}\ee
This entanglement-power relationship characterizes one-dimensional transmission of entropy or information for non-thermal radiation. At thermal equilibrium, this can be compared to Pendry's maximum entropy rate for power, $\dot{S} = \sqrt{\pi P/3}$, also called the noiseless quantum channel capacity, investigated by Bekenstein-Mayo in the context of black holes information flow as (1+1)-dimensional \cite{Bekenstein:2001tj}.  The factor of 2 is accounted for by a uni-directional flow to a single observer, conventionally taken to be situated at $\mathscr{I}^+_R$ future null right infinity. 
The result Eq.~(\ref{EP}) compliments the celebrated Bianchi-Smerlak formula \cite{Bianchi:2014qua} revealing the energy-entanglement connection,
\be 2\pi \mathcal{F}(u) = 6 S'(u)^2 + S''(u), \label{FS}\ee
which implies that the outgoing flux is completely determined by the structure of entanglement at
future null infinity, and vice versa, the entanglement entropy is completely determined by the flux through the second order differential equation from the values of $S(u)$ and $S'(u)$ at a single point at $\mathscr{I}^+_R$.  Eq.~(\ref{FS}) involves derivatives  with respect to retarded time $u$ rather than proper time $\tau$ as is the case in Eq.~(\ref{EP}).

\subsection{Averaging Radiation Reaction}\label{LarmortoLAD}
Turning our attention to the self-force in this subsection, we derive the formula for radiation reaction from our previously derived mirror Larmor power, Eq.~(\ref{Larmor_power}), using an average over proper time.  Our results in this subsection confirm those of Ford-Vilenkin \cite{Ford:1982ct}, building confidence in the overall theme by connecting the self-force to the result for Larmor power. 

Energy lost by radiation for an accelerating point charge tends to slow it down.   This is because there is a backreaction of the radiation on the particle itself.  In the case of a moving mirror this is in practise not so, because the trajectory is usually just assumed apriori.  Allow us to try to find this radiation reaction on the mirror without assuming locally any trajectory and by use of an averaging over proper time.  Starting with the Larmor mirror power, Eq.~(\ref{Larmor_power}), expressed in rapidity Eq.~(\ref{rp}),
\be P = \frac{\eta'(\tau)^2}{6\pi},\ee
the reaction must be, averaged over proper time, the work done on the mirror equal to the negative of the energy lost to the vacuum radiation:
\be \overline{F \eta} = - \frac{\overline{\eta'(\tau)^2}}{6\pi}.\ee
We are still in natural units, $c=1$, and so $\overline{F \eta}$ is an average radiation reaction power linearly and proportionally dependent on the rapidity, $\eta$, of the mirror.  
Writing the proper acceleration as,
\be \eta'(\tau)^2 = \frac{d}{d\tau}(\eta\eta') - \eta\eta'',\ee
where the total derivative with respect to proper time vanishes due to our averaging procedure (or in the case of assuming global asymptotic inertial trajectory\footnote{This is also equivalently accomplished by assuming periodicity.}), we then have
\be \overline{F \eta} = + \frac{\overline{\eta \eta''(\tau)}}{6\pi}.\ee
This allows us to identify the radiation reaction force as,
\be F = \frac{\eta''(\tau)}{6\pi}.\label{force(rapidity)} \ee 
We will show that this result is in agreement with the magnitude, Eq.~(\ref{mirror_LAD1}), of the covariant LAD 4-force Eq.~(\ref{4-vector}), derived using the Davies-Fulling stress tensor, in the next subsection.

\subsection{Confirmation of LAD magnitude}\label{LAD}

A rigorous non-averaging derivation of Eq.~(\ref{force(rapidity)}) is accomplished by relativistic covariance.  In this subsection we derive the mirror LAD force using electromagnetic 4-vector formulation and quantities known in special
relativity, e.g. proper acceleration, celerity and rapidity. 

Before turning to the derivation of the force, let us briefly review some known formulations. The point charge in SI units to moving mirror natural units ($\mu_0 = \epsilon_0 = 1$) has a coupling which can be expressed by the substitution:
\be \frac{2}{3}\left(\frac{q^2}{4\pi \epsilon_0 c^3}\right) = \frac{q^2}{6\pi\epsilon_0 c^3} = \frac{\mu_0 q^2}{6\pi c}\quad \Rightarrow \quad \frac{1}{6\pi}.\label{coupling}\ee
Notice Gaussian units are $4\pi\epsilon_0 = 1$ and $\mu_0 = 4\pi$.
We will need the proper acceleration, $\alpha$, that is a Lorentz invariant, defined by:
\be\alpha^2 \equiv -\frac{d^2x^\mu}{d\tau^2}\frac{d^2x_\mu}{d\tau^2}.\ee
It will also be helpful to have the velocity $v$, the Lorentz factor $\gamma$, and the celerity $w$, which are defined through the rapidity $\eta$,
\be v =  \tanh \eta, \quad \gamma =  \cosh\eta, \quad  w = \sinh\eta, \ee
or similarly,
\be v = \frac{dx}{dt}, \qquad \gamma = \frac{dt}{d\tau}, \qquad  w = \frac{dx}{d\tau}. \ee
The self-force scalar invariant, $F$, can be written via 4-vectors as,
\be F^2 \equiv -F^\mu F_\mu,\label{scalar invariant} \ee
which will be the final object that we obtain in this subsection.

Let us now confirm the mirror self-force by substituting Eqs. (\ref{coupling})-({\ref{scalar invariant}}) into the radiative force introduced by Ford-Vilenkin \cite{Ford:1982ct},
\be 6\pi F^\mu = \frac{d^3 x^\mu}{d\tau^3}-\alpha^2 \frac{dx^\mu}{d\tau}. \label{4-vector} \ee
Here the 4-vector has time $F^0 = F^t$ and space $F^1 = F^x$ components respectively for $6\pi F^\mu$:
\be \gamma''(\tau) -\alpha^2 \gamma = \alpha'(\tau)w,\ee
\be w''(\tau)-\alpha^2 w = \alpha'(\tau) \gamma,\ee
where $\alpha=\eta'(\tau)$.
So, the time and space components are expressed in rapidity as:
\be 6\pi F^t = \eta''(\tau) \sinh\eta, \ee
\be 6\pi F^x = \eta''(\tau) \cosh\eta.\ee
With signature $(+,-,-,-)$ or just $(+,-)$ for our (1+1)-dimensional context, the magnitude can be found by, 
\bea F &=& \sqrt{-F^\mu F_\mu} \\
&=& \sqrt{-(|F^0|^2-|F^1|^2)}\\
 &=& \sqrt{|F^x|^2-|F^t|^2} \\
&=& \frac{\eta''}{6\pi} \sqrt{\cosh^2\eta -\sinh^2\eta}\\
&=& \frac{1}{6\pi} \eta''(\tau),\eea
giving us the simple relationship for jerk,
\be \alpha'(\tau) = 6\pi F,\label{mirror_LAD1}
\ee
which coincides with the Eq.~(\ref{force(rapidity)}) taking into account the change of rapidity with respect to proper time, $\alpha = \eta'(\tau)$.

\subsection{Derivation of LAD formula}\label{mirror_LAD}

Let us now move to the derivation of the self-force from the moving mirror point of view explicitly.
Consider the total energy-momentum emitted to the right of the mirror,
\be E^R = \int_{-\infty}^{\infty} \mathcal{F}^R \d u,\ee
where the Schwarzian derivative defines the quantum stress tensor \cite{Davies:1976hi},
\be \mathcal{F}^R(u) = -\frac{1}{24\pi} \{p(u),u\},\ee
which we convert to proper time \cite{Good:2017ddq}, 
\be \mathcal{F}^R(\tau) = -\frac{1}{12\pi} \eta''(\tau)e^{+2\eta(\tau)},\label{ftau}\ee
so that with Jacobian $du = e^{-\eta} d\tau$, 
\bea E^R &=& -\frac{1}{12\pi} \int_{-\infty}^{\infty} \eta''e^{+2\eta} (e^{-\eta} \d \tau),\\
&=& -\frac{1}{12\pi} \int_{-\infty}^{\infty} \eta''e^{+\eta} \d \tau. \eea
Similarly, with a parity flip, the energy-momentum emitted to left of the mirror is,
\be E^L = +\frac{1}{12\pi} \int_{-\infty}^{\infty} \eta''e^{-\eta} \d \tau. \ee
We are looking for the difference in energy-momentum between the left and right sides of the mirror to construct the 4-vector radiation reaction self-force.  The time component, $F^t = F w$, is constructed from the energy,
\be \Delta U = \int F \d x = \int F v \d t = \int F v \gamma \d \tau = \int F w \d \tau.\ee
On the other hand, the energy is defined as,
\be \Delta U = U^L-U^R = \int_{-\infty}^{\infty} \frac{d U}{d \tau} \d \tau = \int_{-\infty}^{\infty} F^t \d \tau, \ee
where 
\be F^t = \frac{d U}{d \tau} = \frac{d U}{dx} \frac{dx}{d\tau} = F \sinh \eta = F w.\ee
The space component of the force, $F^x = F \gamma$, is constructed from the momentum,
\be \Delta \mathcal{P} = \int F \d t = \int F \gamma \d \tau.\ee
The difference in momentum radiated between the two sides is expressed as,
\be \Delta \mathcal{P} = \mathcal{P}^L-\mathcal{P}^R = \int_{-\infty}^{\infty} \frac{d \mathcal{P}}{d \tau} \d \tau = \int_{-\infty}^{\infty} F^x \d \tau, \ee
which defines the space component piece of the radiation reaction on the mirror.  This is explicitly,
\be F^x = \frac{\eta''}{12\pi}[e^{-\eta}-(-e^{+\eta})]= \frac{\eta''}{6\pi} \cosh \eta = F \gamma,
\ee
where one can already see that $F = \eta''/6\pi$.  
The radiation reaction force,
\be F^\mu = (F^t, F^x) = \gamma( F v, F ) = (w F, \gamma F),\ee
or, equivalently written in covariant notation as in Eq. (\ref{4-vector}),
has a Lorentz scalar invariant jerk,
\be F = \frac{\eta''(\tau)}{6\pi}. \label{self-force}\ee
While our derivation in this subsection for the radiation-reaction force ostensibly relies on the expression for finite conserved energy, the integration is not explicitly taken.  For finite energy one requires the acceleration to vanish asymptotically.  It is safe to assume the self-force does not actually require this constraint.  This is congruent with the usual LAD expression in electrodynamics which holds even in the mathematical case where a charged point is accelerated in both asymptotic limits. 

We have derived the moving mirror LAD formula for the radiation reaction using conservation of energy (the difference in energy-momentum between the right and left sides of the mirror), but made no effort to identify the mechanism responsible for the force.  In the case of a point charge, one imagines the force as the recoil effect of the particle's own field acting back on the charge, but in the case of a mirror we see it is not the source of a field whatsoever.  

In the electromagnetic case, one has the problem of the field blowing up right at the point charge.  But in the mirror model we know the field is identically zero at the mirror.  So what then, is the mechanism? The accepted answer in the case of charge is that an extended charge distribution divided into infinitesimal pieces gives rise to a net force of the charge on itself - the self-force -  as a consequence of the breakdown of Newton's third law within the structure of the particle.  Perhaps a distributed boundary condition calculation could also give rise to a self-force on the mirror.  In other words, the net force exerted by the scalar field generated by different pieces of the distributed boundary condition acts on each other to produce a mirror self-force.  Such a calculation is beyond the scope of this paper.

\subsection{Entanglement \& Radiative Force}\label{entanglementforce}

In the relativistic entanglement-power Section \ref{entanglementpower}, we applied the entanglement-rapidity relationship to the power.  This result gave us insight into the information flow distributed by the radiation.  In this subsection we apply the entanglement-rapidity relationship to the radiation reaction itself. We find the self-force is proportional to the second derivative of the von Neumann entanglement entropy in a simple entanglement-force relationship.

Using the Davies-Fulling exact relativistic quantum stress tensor, expressed in proper time $\tau$ as $T_{00} = \mathcal{F}(\tau)$ \cite{Good:2017ddq} where,
\be     12\pi \mathcal{F}(\tau) = -\eta''(\tau)e^{2\eta(\tau)},\ee
as well as the  Lorentz-Abraham-Dirac 4-force in one dimension, we have demonstrated  that $F^t = F w$, where the celerity is $w=dx/d\tau$, and $F^x = F \gamma$, where the Lorentz factor is $\gamma = dt/d\tau$, (see the closely related results of Higuchi-Martin \cite{Higuchi:2005gh}). These results ultimately led to,
\be 6\pi F(\tau) = \alpha'(\tau). \label{jerk}\ee
The derivation reveals Eq.~(\ref{jerk}) as the most simple interpretation of the LAD force, with magnitude $F$ of Eq.~(\ref{4-vector}) as the jerk of the mirror, i.e. the proper time derivative of the proper acceleration. The source of this force is the reaction of the scalar field to the presence of the accelerating mirror in vacuum.  The proper time derivative of the proper acceleration determines the force and can be nonzero even when the acceleration itself of the mirror is instantaneously zero, and the mirror is not radiating particles. The disturbing implications of the Lorentz-Abraham-Dirac formula which are still not entirely understood in classical electrodynamics (see e.g. \cite{Rohrlich:1997,Yaghjian:2006}), carry over in analog, to the quantum scalar field of the moving mirror model.

Using the rapidity-entanglement relationship, $\eta = -6 S$ in Eq. (\ref{self-force}), it is easy to find 
the von Nuemann entanglement entropy in terms of the radiative reaction force,
\be S''(\tau) = -\pi F(\tau).\label{entropy-force}\ee
This relationship connects information flow in the system to the self-force on the mirror.  In a similar vein to the interesting features like negativity and thermodynamic interpretations of entropic forces \cite{Visser:2011jp}, this entanglement self-force also assumes negative values and demonstrates an information interpretation of the radiative reaction force: it is the second proper time derivative of the von Neumann entanglement entropy.\footnote{The sign in $S'' = -\pi F$ tells us that when the force on the mirror is to the left, away from the observer at $\mathscr{I}^+_R$, then $S''$ is positive. The sign is by convention because the observer is chosen to be located on the right.}

It would be interesting to know whether Eq.~(\ref{entropy-force}) holds outside the moving mirror model considering the closely related accelerated boundary correspondences (ABCs) with cosmologies and black holes. Regardless, advances in general relativity, like the maximum force conjecture, may play a role in better understanding the moving mirror model and associated entanglement entropy.  There have been a number of works in the last decades indicating that $F_{\max} = c^4/(4G)$ is the limiting force (see \cite{Barrow:2014cga, Good_2015Springlike} for more references) in general relativity. Taking into account entropy-force relation of Eq.~(\ref{entropy-force}), this implies a constraint on the rate of change of the entanglement entropy, i.e. if the force, or jerk $\alpha'(\tau)$, has this maximum, then the second derivative of entropy has a minimum possible value, $S''_{\textrm{min}} = -\pi/4$.

\section{CGHS Larmor power and self-force}\label{sec:CGHS power & force}

Having derived the power and self-force for any moving mirror in general, we now specialize to the exponentially accelerated mirror that has particle production which corresponds to the CGHS system. We apply the Larmor power and LAD force derived in previous sections to our particular CGHS mirror and find that a simple entanglement-over-distance relationship is revealed connecting the entanglement-rapidity relationship to the space traversed.  In addition, the loss of unitarity is explicitly manifest in the divergence of the power and self-force at the time the horizon forms in the proper frame.

Let us start from the trajectory of the CGHS mirror in spacetime coordinates \cite{Good:2017ddq},
\begin{equation}
z(t)=-\frac{1}{\kappa}\sinh^{-1}\left(\frac{e^{\kappa (t-v_H)}}{2}\right).\label{Arcx_trajectory}
\end{equation}

\begin{figure}[h]
   \centering
   \includegraphics[width=0.9\linewidth]{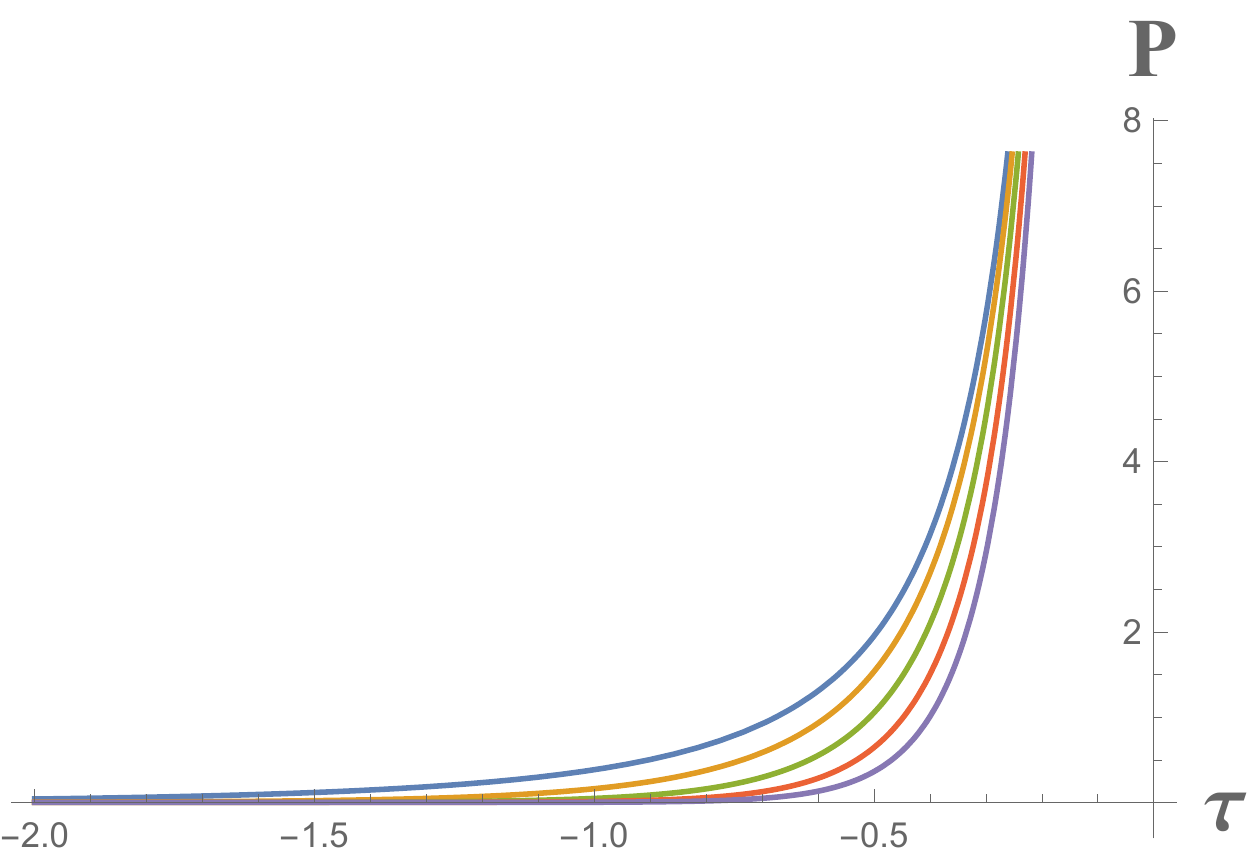}
    \caption{Larmor power for CGHS mirror, Eq. (\ref{power1}), where plots from left to the right correspond to $\kappa=1,2,3,4,5$ cases, respectively, $\hbar=c=1$, and power is normalised by $10$. The power increases asymptotically as time approaches $\tau=0$. The key takeaway is this divergence at a finite proper time when the horizon forms.} 
    \label{fig:Power}
\end{figure}

\begin{figure}[h]
  \centering
   \includegraphics[width=0.9\linewidth]{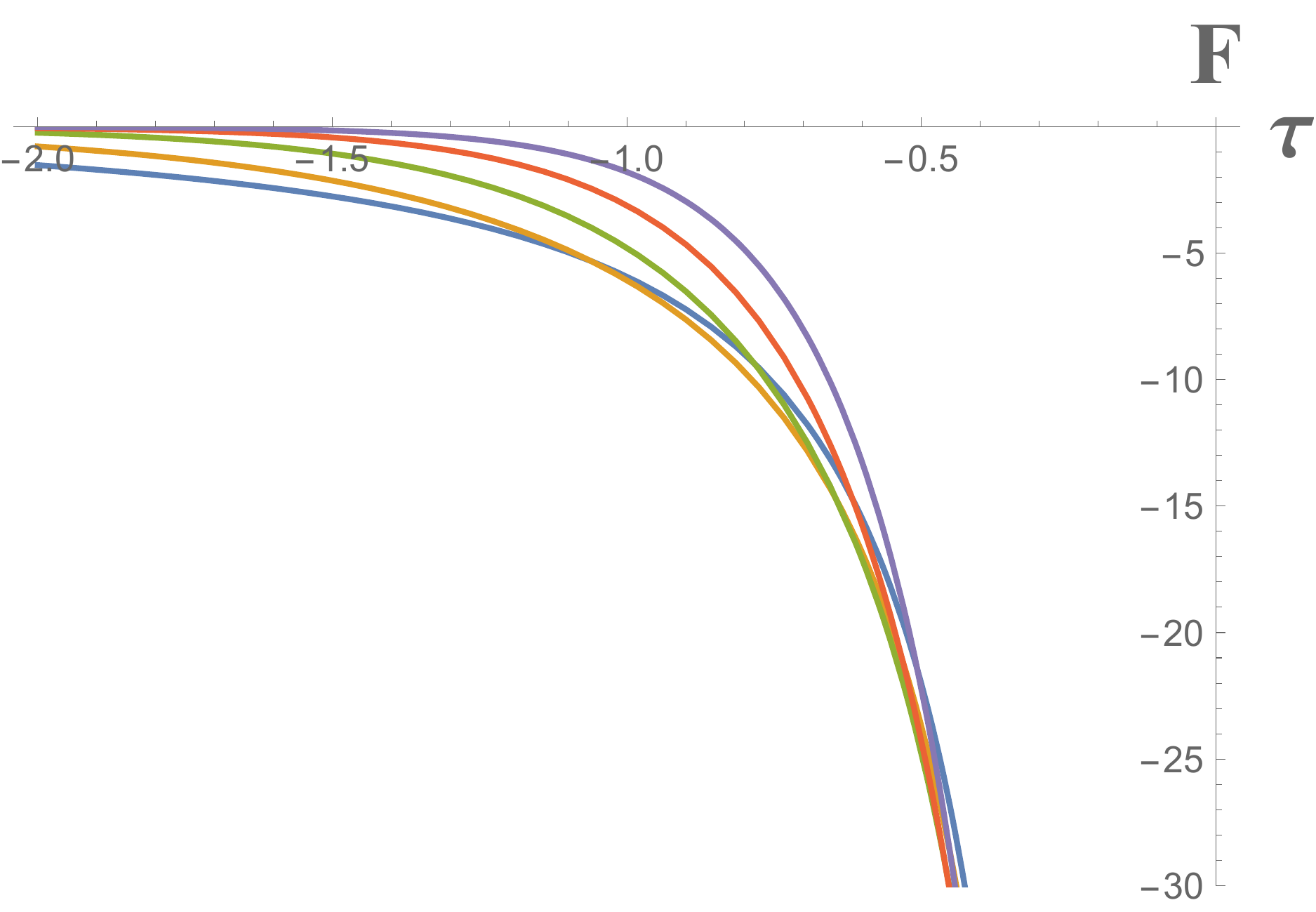}
    \caption{Radiative reaction or the self-force for CGHS mirror, Eq. (\ref{force}), where blue, orange, green, red and purple lines correspond to $\kappa=1,2,3,4,5$ cases, respectively, $\hbar=c=1$, and force is normalised by $10^2$. The self-force has opposite direction with respect to the Larmor power and demonstrates left-hand side trend as the mirror is moving to the left.} 
    \label{fig:Force}
\end{figure}

\noindent The Larmor power and self-force for the CGHS mirror are found using Eq. (\ref{Larmor_power}) and Eq. (\ref{mirror_LAD1}), where $\alpha(\tau)$ is the acceleration in proper time. The procedure of defining and deriving $\alpha(\tau)$ is given in \cite{Good:2017ddq}. Let us start from the connection between proper and coordinate times. For CGHS mirror it is obtained to be,
\begin{equation}
\tau(t)=\int\frac{dt}{\gamma(t)}=\frac{1}{2\kappa}\ln\left|\frac{\sqrt{4+e^{2\kappa (t-v_H)}}-2}{\sqrt{4+e^{2\kappa (t-v_H)}}+2}\right|.\label{proper_time}    
\end{equation} The inverse of Eq. (\ref{proper_time}) yields,
\begin{equation}
t(\tau)=v_H+\frac{1}{2\kappa}\ln\left|\frac{16e^{2\kappa\tau}}{(1-e^{2\kappa\tau})^2}\right|.
\end{equation}Applying it into Eq. (\ref{Arcx_trajectory}) leads to the trajectory in proper time,
\begin{equation}
z(\tau)=\frac{1}{\kappa}\sinh^{-1}\left(\csch(\kappa\tau)\right).
\end{equation}
The next step to obtain $\alpha(\tau)$ is to find celerity and then rapidity. 
Using the rapidity, the proper acceleration is found to be,
\begin{equation}
\alpha(\tau)=\frac{d\eta(\tau)}{d\tau}=\kappa\csch(\kappa\tau). \label{proper_acceleration} 
\end{equation}
This result, Eq.~(\ref{proper_acceleration}), is found using a different method in Ju\'arez-Aubry \cite{Benito:2017} and is in agreement. 
Substituting Eq. (\ref{proper_acceleration}) into Eq. (\ref{Larmor_power}) and Eq. (\ref{mirror_LAD1}), we obtain corresponding Larmor power and radiation reaction force for the CGHS mirror as, 
\begin{equation}
P=\frac{\alpha^2}{6\pi}=\frac{\hbar}{c^2}\frac{ \kappa^2\csch^2(\frac{\kappa\tau}{c})}{6\pi},  \label{power1} 
\end{equation}
and
\begin{equation}
F=\frac{\alpha'}{6\pi}=-\frac{\hbar}{c^3}\frac{\kappa^2}{6\pi}\coth(\frac{\kappa\tau}{c})\csch(\frac{\kappa\tau}{c}). \label{force}   
\end{equation} 
The terms on the right of Eq. (\ref{power1}) and Eq. (\ref{force}) have reinstated $\hbar$ and $c$, noting that $\kappa$ has units of an acceleration, in order to emphasize they are a quantum Larmor power and quantum self-force, respectively. The dependences of the CGHS mirror Larmor power and self-force on proper time and $\kappa$, Eq. (\ref{power1}) and Eq. (\ref{force}), are demonstrated graphically in Figs.~\ref{fig:Power} and \ref{fig:Force}.

Fig.~\ref{fig:Force} has lines corresponding to different values of $\kappa$ which intersect. This is explained by the fact that the dependence of the CGHS self-force, Eq.~(\ref{force}), on the single parameter of the system $\kappa$ is non-trivially different from the dependence of the power, Eq.~(\ref{power1}). 

Let us now consider the timespace trajectory of the mirror in natural units,
\begin{equation}
t(x)=v_H+\frac{1}{\kappa}\ln[-2\sinh(\kappa x)].   
\end{equation}Using this form of the trajectory we find rapidity in terms of space coordinate $x$, 
\begin{equation}
\eta(x)=\kappa x.\label{kx}    
\end{equation}
So, the rapidity, or the information defining dynamical quantity, in terms of $x$ has surprisingly simple form: it linearly depends on the space coordinate.  Eq.~(\ref{kx}) is the simplest way to express the trajectory of the CGHS mirror.   

\begin{figure}[h]
  \centering
   \includegraphics[width=0.9\linewidth]{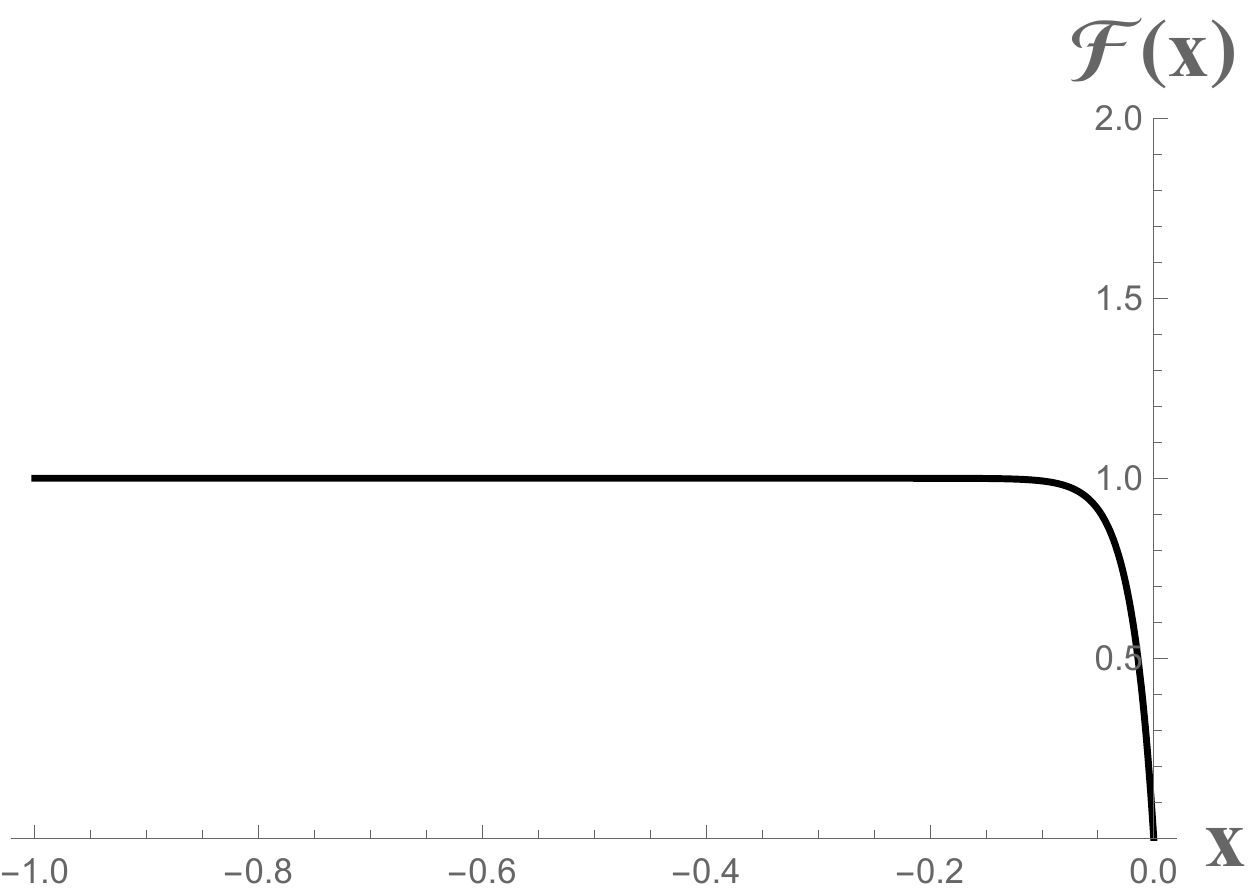}
    \caption{This graph should be read from right to left mapping the initial motion of the mirror in the past to the future. This is the energy flux for the CGHS mirror, Eq.~(\ref{EnF}), with $\kappa=\sqrt{48\pi}$ for illustration so that at thermality the flux is equal to one. The mirror moves to the left, while the flux ascends from $x=0$, then monotonically approaches constant thermal emission at late positions, $x=-\infty$.} 
    \label{fig:flux}
\end{figure}

The last interesting quantity we compute is the energy flux in terms of $x$. Using Eq. (\ref{flux(x)}), the CGHS mirror flux is found to be,
\begin{equation}
\mathcal{F}(x) = \frac{\kappa^2}{48\pi}(1-e^{4\kappa x}).\label{EnF} 
\end{equation} This form immediately clarifies that at late times (far-left positions, $x\to-\infty$), the energy flux is a constant associated with thermal emission that is in agreement with the thermal behaviour of the CGHS black hole radiation, $\mathcal{F} = \kappa^2/(48\pi)$. The graphical illustration of this flux is shown in Fig. \ref{fig:flux}.

\section{Conclusions \& Future Work}\label{sec:summary}

Overall, the equivalence between the CGHS black hole and the exponentially accelerating moving mirror in lab time can be seen from several explicit matching quantities: the matching condition for the CGHS black hole and the trajectory of the mirror in null coordinates, the spectra, and consequently, the temperatures. A one-to-one correspondence is ensured as long as $|\Lambda|=|\kappa|$ requirement is met. The correspondence is summarized in Table \ref{III}. 

An important physical result of this correspondence is the demonstration that the moving mirror horizon corresponds in the black hole system to a quantity determined by the black hole parameters $\Lambda$ and $M$. This is understood by seeing that the horizon location of the mirror is associated with the CGHS black hole spacetime geometry as determined by the global non-zero curvature.  This geometry, in turn, is determined by the cosmological constant and curvature caused by the black hole mass. Interestingly, it has been found that unlike Schwarzschild black hole case, where the singularity and the event horizon are located at different positions, the CGHS mirror mimics both the event horizon and the center of the CGHS black hole as they happen at the same location, $r=0$.  This overlap is almost assuredly responsible for the particular simplicity of the mathematics in the mirror case and the ease and utility in describing this exact spectrum via a simple accelerating trajectory in a flat-spacetime background.  It also highlights no conflict between the origin singularity and event horizon as the location of particle production, since they are both one and the same.

More general considerations have given us the Larmor power radiated by an arbitrary moving mirror and the LAD formula for the radiation reaction.  The derivations utilize general dynamics of the mirror expressed in terms of proper acceleration and rapidity, and lead naturally to an information interpretation by expressing the rapidity in terms of entanglement entropy. The power and force are found to have the same dynamic form as that in classical electrodynamics for a moving point charge.  In terms of information, the entanglement power and entanglement self-force are interpreted in terms of first and second derivatives of the von-Neuman entanglement entropy, respectively.

Specializing to our particular CGHS moving mirror, the Larmor power is found to diverge as $\tau\to 0^-$. As proper time ticks to $\tau=0$, the mirror is infinitely accelerating, reaching the speed of light. Consistently, the direction of the self-force is opposite the direction of the radiated Larmor power. It is worth emphasizing that as a guide, in SI units both the Larmor power and LAD force for the CGHS mirror are proportional to $\hbar$, underscoring the fact that the power and self-force are quantum (not classical) measures.

Lastly, the CGHS mirror has two simplifying results when expressed in terms of space rather than time or light-cone coordinates:  the trajectory rapidity is simply proportional to the distance travelled, $\eta = \kappa x$ and the radiative flux emitted by the CGHS mirror is seen by eye as thermal (constant emission) at far-left positions (late times).  In summary list-form, the salient features of this work are:
\begin{itemize}
 \item CGHS mirror $\Leftrightarrow$ EH \& BH center; \ref{sec:expomirror}
\item Mirror Larmor power; $P\sim\alpha^2(\tau)$; \ref{power}
    \item Entanglement-power; $ P \sim S'(\tau)^2$; \ref{entanglementpower}
      \item Larmor to LAD Averaging; $P\rightarrow F$; \ref{LarmortoLAD}
    \item Mirror LAD self-force; $ F \sim \alpha'(\tau)$; \ref{LAD} \& \ref{mirror_LAD} 
   \item Entanglement-force; $ F \sim S''(\tau)$; \ref{entanglementforce}
   
    \item CGHS self-force and CGHS power; \ref{sec:CGHS power & force}
    \end{itemize}

Future extensions of this work are foreseen.  Hawking \cite{Hawking:1974sw} pointed out that at very early times of gravitational collapse, a star cannot be described by the no-hair theorem. So in this context, a variety of different collapse situations corresponds to different mirror trajectories.  It is likely to be fruitful to consider modifications to the mirror trajectory (one of which has already been done to CGHS e.g. \cite{good2013time}) made to provide different
early time approaches to a thermal distribution, particularly those modifications that
can afford unitarity and finite evaporation energy, modeling more realistic situations congruent with finite mass black holes and quantum purity.

The modifications that can take into
account energy conservation like those of the dilaton gravity models and moving mirror models have had significant
success as a laboratory for studying black hole evaporation. The physical problem in (1+1)
dilaton gravity of the evaporating black hole and its modified emission extends to complete
evaporation for the Russo, Susskind, and Thorlacius (RST) model \cite{Russo:1992ax} and to partial
evaporation leaving a remnant for the Bose, Parker, and Peleg (BPP) model \cite{Bose:1995pz}. The two-dimensional RST model for evaporating black holes is locally equivalent - at the full quantum level - to Jackiw-Teitelboim (JT) gravity that was recently shown to be unitary \cite{Fitkevich:2020okl}.

The similarity of drifting moving mirrors (see e.g. \cite{Good:2016atu}) to the BPP model is striking in several qualitative aspects:
NEF emission as a thunderpop (NEF emission from evaporating black holes, at least in the case of a (1+1)-dimensional dilaton gravity, has already been known in the literature for over 20 years \cite{Bose:1995bk}), a left over remnant, and finite total energy emission. It
is also interesting that the mass of the remnant in the BPP model is independent of the
mass $M$ of the infalling matter, since with respect to the issue of energy conservation,
there is no known physical analog for $M = 1/(4 \kappa)$, the initial mass of the shockwave, in
the mirror model.

We hope this work offers insight for future direction, using the formulas for mirror radiative power and radiation reaction force.  Investigations of behavior of the Larmor power and LAD force of other existing moving mirror models will be used to compare results to better understand the specific physics of moving mirrors and the general physics of acceleration radiation from the quantum vacuum.\\


\onecolumngrid

\begin{table}[h!]
\caption{Matching quantities for CGHS black hole and exponentially accelerating mirror.}
\centering
 \begin{tabular}{c| c| c} 
 \hline
 Quantity & black hole & moving mirror \\
 \hline\hline
 Trajectory &  $u(U)=U-\frac{1}{\Lambda}\ln\left(1-e^{\Lambda(U-v_0)}\right)$ & $f(v)=v-\frac{1}{\kappa}\ln\left(1-e^{\kappa(v-v_H)}\right)$   \\ 
 [2.5ex]
 Spectrum & $|\beta_{\w\w'}|^2=\frac{1}{4\pi^2\Lambda^2}\frac{\omega'}{\omega}\left|B\left[\frac{i\omega_+}{\Lambda},1-\frac{i\omega}{\Lambda}\right]\right|^2$ & $|\beta_{\w\w'}|^2=\frac{1}{4\pi^2\kappa^2}\frac{\omega'}{\omega}\left|B\left[\frac{i\omega_+}{\kappa},1-\frac{i\omega}{\kappa}\right]\right|^2$  \\
 [2.5ex]
 Temperature & $T = \frac{\Lambda}{2\pi}$ & $T = \frac{\kappa}{2\pi}$   \\
 [1.0ex]
 \hline
 \end{tabular}
 \label{III}
\end{table}
\twocolumngrid


\acknowledgments 
Funding from state-targeted program ``Center of Excellence for Fundamental and Applied Physics" (BR05236454) by the Ministry of Education and Science of the Republic of Kazakhstan is acknowledged. MG is also funded by the FY2018-SGP-1-STMM Faculty Development Competitive Research Grant No. 090118FD5350 at Nazarbaev University.

\bibliography{main} 

\end{document}